\documentclass[twoside]{article}
\usepackage{Style.V_7}

\title{Solar System and Atomic Clock Bounds on Locally Coupled Swampland Scalars}
\paperidentifier{Published in \textit{Nuclear Physics B} [10.1016/j.nuclphysb.2026.117629].}
\paperauthor[1]{Suraj Gavhale}
\paperauthor[1]{Maxim Khlopov}
\paperauthor[2]{Oem Trivedi}
\paperauthor[3,4]{Maxim Krasnov}
\paperaffiliation[1]{Virtual Institute of Astroparticle Physics, Paris 75018, France}
\paperaffiliation[2]{Department of Physics and Astronomy, Vanderbilt University, Nashville 37235, USA}
\paperaffiliation[3]{Research Institute of Physics, Southern Federal University, Rostov-on-Don 344090, Russia}
\paperaffiliation[4]{National Research Nuclear University MEPhI, Moscow 115409, Russia}

\newcommand{\Mpl}{M_{\rm Pl}}
\newcommand{\Ode}{\Omega_{{\rm DE},0}}
\newcommand{\Om}{\Omega_{m,0}}
\newcommand{\Ob}{\Omega_{b,0}}
\newcommand{\Oc}{\Omega_{c,0}}
\newcommand{\dd}{\mathrm{d}}

\newcommand{\cstar}{c_{\rm dS}}
\newcommand{\etastar}{c'_{\rm dS}}

\newcommand{\dvec}{\mathbf d}

\begin{document}
\maketitle

\begin{abstract}
We study how local measurements constrain light scalar fields that are relevant in late time cosmic acceleration and are often discussed in connection with swampland criteria. Starting from a scalar-tensor framework, we define the swampland slope and curvature variables in the canonically normalized Einstein frame and relate them to Solar System tracking, Lunar Laser Ranging, equivalence principle tests and atomic clock comparisons. These measurements do not constrain the scalar velocity by itself, but products of that velocity with microscopic couplings to matter, gravity and atomic parameters. For coupling directions visible to local experiments, the scalar is driven into an ultra slow present regime. This severely restricts the possibility of realizing an $\mathcal{O}(1)$ de Sitter gradient through unscreened visible sector couplings alone. The refined de Sitter alternative remains possible, but in the single field hilltop realization it requires proximity to the maximum or a tuned suppression of the growing mode.
\end{abstract}

\vfill
\paragraph{Acknowledgment.} The research of M.K. was carried out in the Southern Federal University with financial support from the Ministry of Science and Higher Education of the Russian Federation (State contract FENW-2026-0028). The work of O.T. is supported in part by the Vanderbilt Discovery Doctoral Fellowship.

\newpage

\section{Introduction}

The swampland program begins from a simple fact, whether a low energy effective field theory can be consistent and still fail to arise as the infrared limit of quantum gravity. This viewpoint goes back to Vafa's original proposal and has since been developed in broad reviews of quantum gravity constraints on effective theories \cite{Vafa2005,BrennanCartaVafa2017,Palti2019}. For scalar sectors, two frequently discussed statements are the distance conjecture and the de Sitter conjecture. The distance conjecture associates a parametrically large geodesic distance in field space with an infinite tower of states becoming light, and therefore with the loss of validity of the original finite field EFT \cite{OoguriVafa2006,Palti2019}. For a canonically normalized field, define the dimensionless local slope $\lambda_V$ and the smallest dimensionless Hessian eigenvalue $\eta_V$. The original de Sitter proposal is
\begin{equation}
\lambda_V\equiv \Mpl\frac{|\nabla V|}{V}\geq \cstar
\label{eq:dS-gradient}
\end{equation}
where $\cstar=O(1)$ is a conjectural threshold rather than a new name for $\lambda_V$ \cite{Obied2018,AgrawalObiedSteinhardtVafa2018,RaveriHuSethi2018}. Refined versions permit the slope condition to fail when
\begin{equation}
\eta_V\equiv \Mpl^2\frac{\operatorname{eig}_{\min}(\nabla_i\nabla_jV)}{V} \leq-\etastar ,
\label{eq:refined-dS}
\end{equation}
with $\etastar=O(1)$ \cite{GargKrishnan2018,OoguriPaltiShiuVafa2018}. Thus $\lambda_V$ and $\eta_V$ are field dependent, while $\cstar$ and $\etastar$ are the conjectural order one constants against which they are compared. The strength of the evidence depends on the region of field space, the degree of parametric control and the version of the conjecture being used. 

In particular, arguments derived in asymptotic, weak coupling limits need not determine the same coefficient at a generic finite distance point. The status of de Sitter constructions, the scope of the conjecture and several proposed refinements are reviewed or discussed in ~\cite{DanielssonVanRiet2018,Palti2019,Andriot2018,HebeckerWrase2019,AndriotRoupec2019}. We therefore retain $\cstar$ and $\etastar$ and show representative consequences for more than one numerical choice rather than identifying ``order one'' with exactly unity. Our question is deliberately local. We ask what part of this parameter space is probed when a late time scalar is light, coupled to visible matter or to the effective gravitational coupling, and unscreened on laboratory and Solar System scales.

Such a field is not only a cosmological background and it also enters Lunar Laser Ranging (LLR), post-Newtonian tracking, weak equivalence principle searches and atomic clock comparisons \cite{ac1Marion:2002iw,ac2Barontini:2021mvu,ac3Sherrill:2023zah,ac4Flambaum:2007my,ac5godun2014frequency}. Trivedi's analyses of swampland constraints and atomic clock scalar-tensor dark energy provide closely related motivation \cite{Trivedi2020,Trivedi2025}. Clock comparisons constrain only the microscopic coupling directions to which the chosen transitions are sensitive. LLR constrains the time variation of the locally measured gravitational coupling, not the scalar velocity alone. Post-Newtonian and equivalence principle measurements constrain source dependent and composition dependent scalar charges \cite{ppn4Hou:2017cjy,ppn2Dzuba:2024pri,ppn3Dent:2008ev,ppn5Zhang:2023nil,ppn6Saito:2024xdx}. Screening, hidden sector dominated couplings and locally invisible coupling directions can therefore evade the simplest inference \cite{sc1burrage2014screening,sc2fischer2024screened,sc3shibata2023properties,sc4chakrabarti2022screening,sc5nakamura2021dynamical}. We keep this conditionality in mind. In terms of the dimensionless velocity of the present cosmological scalar
\begin{equation*}
x= \frac{\dot\phi_0}{H_0\Mpl},
\end{equation*}
clock visible directions can lead to the representative ultra slow prior $|x|\lesssim10^{-2}$. By contrast, the current LLR scale is $|\alpha_{M,0}|=O(10^{-4})$ in the approximation relating $\dot G/G$ directly to the Planck mass running. Neither number is universal and the former depends on clock sensitivity coefficients and the latter constrains the product of the scalar velocity with the gravitational coupling direction \cite{ac1Marion:2002iw,ac6Uzan:2024ded,ac7Brzeminski:2022sde,ac8Levy:2024vyd,ac9Arvanitaki:2014faa,ac10Kennedy:2020bac}.

\section{Frames, local observables and swampland variables}\label{sec:local-constraints}

Since swampland quantities $\lambda_V$ and $\eta_V$ are field space quantities, they are best defined using an Einstein frame canonically normalized scalar coordinate. We take the Jordan frame action in the dimensionless $F$ convention
\begin{equation}
        S_J=\int \dd^4x\sqrt{-\tilde g}\left[\frac{\Mpl^2}{2}F(\Phi)\tilde R-\frac12 K_J(\Phi)(\tilde\nabla\Phi)^2-U(\Phi)\right]+S_m[\tilde g_{\mu\nu},\Psi_m;q_a(\Phi)]
\end{equation}
where $F(\Phi_0)=1$ at the present background point and Standard Model parameters $q_a$ may depend on the scalar \cite{BransDicke1961, FujiiMaeda2003, CliftonFerreiraPadillaSkordis2012}. The conformal transformation $g^E_{\mu\nu}=F(\Phi)\tilde g_{\mu\nu}$ gets the gravitational sector to Einstein form\footnote{After the conformal transformation, the coefficient of \((\nabla\Phi)^2\) in the Einstein frame is 
\[\mathcal G^E_{\Phi\Phi}=\frac{K_J(\Phi)}{F(\Phi)}+\frac{3\Mpl^2}{2}\left(\frac{F_{,\Phi}}{F}\right)^2.\] 
and \cref{eq:transformation} defines the canonical coordinate through \(\ (\dd\phi/\dd\Phi)^2=\mathcal G^E_{\Phi\Phi}\). Thus even \(K_J=1\) does not make \(\Phi\) canonical in the Einstein frame when \(F_{,\Phi}\neq0\). The transformed potential is 
\[
V(\phi)=\frac{U(\Phi)}{F^2(\Phi)}.
\]} . The corresponding normalized Einstein frame scalar $\phi$ is defined locally by
\begin{equation}
        \left(\frac{\dd\phi}{\dd\Phi}\right)^2=\frac{K_J(\Phi)}{F(\Phi)}+\frac{3\Mpl^2}{2}\left(\frac{F_{,\Phi}}{F}\right)^2.
        \label{eq:transformation}
\end{equation}
A familiar nontrivial example is the tree level string frame dilaton action
\begin{equation}
S_{\rm s} = \frac{\Mpl^2}{2}\int\dd^4x\sqrt{-\tilde g}\, e^{-2\Phi} \left[ \tilde R+4(\tilde\nabla\Phi)^2+\cdots \right].
 \label{eq:string-frame}
\end{equation}
For a dimensionless dilaton, matching \cref{eq:string-frame} to our convention gives $F(\Phi)=e^{-2\Phi}$ and $K_J(\Phi)=-4\Mpl^2e^{-2\Phi}$. If an overall factor of $\Mpl^2$ is removed from the definition of $K_J$, the second relation is often written simply as $K_J=-4e^{-2\Phi}$ \cite{GasperiniVeneziano2003,FujiiMaeda2003}. \Cref{eq:transformation} gives $(\dd\phi/\dd\Phi)^2=2\Mpl^2$ after the conformal contribution is included. This example also makes the direction of the running transparent. Since $M_*^2=\Mpl^2e^{-2\Phi}$, an increasing dilaton lowers the effective Planck mass and, in the leading $G_{\rm eff}\propto M_*^{-2}$ approximation, strengthens the gravitational coupling. 

All quantities in this paper, including $x$, $y$, $\lambda_V$, $\eta_V$, $d_i$, $\alpha_A$, $\beta_m$, and $\alpha_M$ are understood to be expressed with respect to this normalized Einstein frame field, after which we refer to it simply as $\phi$. With this convention, the one field swampland variables are
\begin{equation}
        \lambda_V=\Mpl\frac{|V_{,\phi}|}{V},
\end{equation}
\begin{equation}
        \eta_V=\Mpl^2\frac{V_{,\phi\phi}}{V}.
\end{equation}
This convention is essential because $V'/V$ and $V''/V$ are not invariant under arbitrary field redefinitions unless the field space metric has been accounted for \footnote{In a multifield formulation, the corresponding quantities are \(M_{\rm Pl}\sqrt{\mathcal G^{ij}\partial_i V\partial_j V}/V\) and the eigenvalues of the covariant Hessian \(M_{\rm Pl}^2\nabla_i\nabla_jV/V\), with \(\mathcal G_{ij}\) the scalar field space metric. The present analysis is the local single canonical field projection of this more general view.}. In the Einstein frame, a simple conformally coupled scalar sector can be written as
\begin{equation}
    S_E=\int\dd^4x\sqrt{-g}\left[\frac{\Mpl^2}{2}R-\frac12(\nabla\phi)^2-(\phi)\right]+S_m[A^2(\phi)g_{\mu\nu},\Psi_m]
        \label{eq:einstein-action}
\end{equation}
for the universal conformal limit. More general dilaton couplings are introduced below through the scalar dependence of the microscopic Standard Model parameters. 

In either description, the scalar entering $\lambda_V$ and $\eta_V$ is the canonically normalized Einstein frame field. For a nearly static spherical source, the two parameters most relevant here appear in the metric as
\begin{equation}
 g_{00}=-1+2U-2\beta_{\rm PPN}U^2+\cdots, 
 \label{eq:ppn-metric}
\end{equation}
\begin{equation}
    g_{ij}=(1+2\gamma_{\rm PPN}U)\delta_{ij}+\cdots ,
\end{equation}
where $U$ is the Newtonian potential and general relativity predicts $\beta_{\rm PPN}=\gamma_{\rm PPN}=1$ \cite{Will2014}. The parameter $\gamma_{\rm PPN}$ measures the spatial curvature produced per unit source mass. Cassini constrained it through the additional light travel time, or Shapiro delay, of radio signals passing near the Sun. 

In a light scalar theory, the same measurement bounds the scalar charge of the Sun. Repeated laser ranging to lunar retroreflectors follows the Earth--Moon orbit over decades. A secular $\dot G/G$ changes the orbital dynamics, while composition dependent Earth and Moon accelerations test the equivalence principle. MICROSCOPE performs a better weak equivalence principle comparison by monitoring the differential acceleration of titanium and platinum test masses in orbit. Atomic clocks ask a different question, and that is, do two transition frequencies, with different dependence on dimensionless Standard Model parameters, drift relative to one another? The four measurements therefore constrain different projections of the same microscopic scalar couplings.

\begin{table}[t]
\centering
\footnotesize
\renewcommand{\arraystretch}{1.4} 
\caption{Representative local inputs used for the numerical scale of the analysis. Errors are reported by the experimental analyses and are not shown in a likelihood here. The final column shows the scalar quantity constrained in this paper and none of these measurements is a model independent bound on $x$.}
\label{tab:local-bounds}
\begin{tabularx}{\textwidth}{@{}
  >{\raggedright\arraybackslash}p{0.16\textwidth}
  >{\raggedright\arraybackslash}p{0.22\textwidth}
  >{\raggedright\arraybackslash}p{0.32\textwidth}
  >{\raggedright\arraybackslash}X @{}}
\toprule
\textbf{Probe} & \textbf{Reported observable} & \textbf{Published result} & \textbf{Scalar interpretation} \\
\midrule
Cassini solar conjunction
& Shapiro delay, expressed as $\gamma_{\mathrm{PPN}}-1$
& $(2.1\pm2.3)\times10^{-5}$ \cite{BertottiIessTortora2003}
& Solar scalar charge, and hence a universal long range coupling \\

LLR
& Secular variation $\dot G/G$
& $(-5.0\pm9.6)\times10^{-15}\ \mathrm{yr}^{-1}$ \cite{BiskupekMullerTorre2021}
& Product $d_Fx=\alpha_{M,0}$, subject to scalar exchange corrections \\

MICROSCOPE
& E\"otv\"os ratio $\eta_{\mathrm{Ti,Pt}}$
& $[-1.5\pm2.3\,(\mathrm{stat})\pm1.5\,(\mathrm{syst})]\times10^{-15}$ \cite{Touboul2022}
& Composition dependent charge difference times the Earth's source charge \\

Atomic clocks
& Linear drift of the fine structure constant
& $\dot\alpha/\alpha=1.8(2.5)\times10^{-19}\ \mathrm{yr}^{-1}$ \cite{Filzinger2023}
& Electromagnetic projection $x\,\Delta k_{E3,E2}(\dvec)$ \\
\bottomrule
\end{tabularx}
\end{table}

The locally inferred Newton coupling scales as $G_{\rm eff}\propto1/F(\phi)$ only at leading order. A Cavendish or PPN interpretation must also include direct scalar exchange and the scalar charges of the source and test bodies \cite{DamourEspositoFarese1992,Will2014}. The effective Planck mass is $M_*^2(\phi)=\Mpl^2F(\phi)$ and its EFT of dark energy running is
\begin{equation}
        \alpha_M\equiv \frac{\dd\ln M_*^2}{\dd\ln a}=\frac{\dd\ln F}{\dd\ln a}=\frac{1}{H}\frac{\dot F}{F}.
        \label{eq:alphaM-def}
\end{equation}
Therefore, in the approximation in which the direct scalar exchange contribution to the measured Newton coupling is either separately accounted for or subleading in the time derivative
\begin{equation}
        \frac{\dot G}{G}=-\frac{\dot F}{F}=-\alpha_M H.
\end{equation}
Using the LLR result in Table~\ref{tab:local-bounds} and $H_0=6.89\times10^{-11}\ {\rm yr}^{-1}$, a conservative Gaussian $95\%$ envelope gives
\begin{equation}
 \epsilon_G^{95}
 \equiv
 \frac{|(\dot G/G)_{\rm obs}|+1.96\,\sigma_G}{H_0}
 \simeq3.5\times10^{-4}.
 \label{eq:epsilonG95}
\end{equation}
Thus $|\alpha_{M,0}|\lesssim\epsilon_G^{95}$ in this approximation. The bound is about two orders of magnitude tighter than the older $O(10^{-2})$ scale, but it remains a constraint on $d_Fx$, not on $x$ alone. 

Atomic clocks constrain the time variation of dimensionless frequency ratios. For a light scalar coupled to the microscopic parameters controlling atomic transitions, a clock transition $i$ has the local expansion
\begin{equation}
        \nu_i(\phi)=\nu_i^{(0)}\left[1+k_i\frac{\phi-\phi_0}{\Mpl}+O\left(\frac{(\phi-\phi_0)^2}{\Mpl^2}\right)\right].
\end{equation}
This linear sensitivity coefficient expansion can be found also in analyses of varying fundamental constants and scalar couplings to atomic transition frequencies \cite{Rosenband2008,Uzan2011,Safronova2018}. For two clock transitions,
\begin{equation}
        D_{21}\equiv \frac{\dd}{\dd t}\ln\left(\frac{\nu_2}{\nu_1}\right)\simeq (k_2-k_1)\frac{\dot\phi_0}{\Mpl}.
\end{equation}
\begin{equation}
        x\equiv\frac{\dot\phi_0}{H_0\Mpl} \simeq \frac{D_{21}}{H_0(k_2-k_1)}.
        \label{eq:x-clock}
\end{equation}
\Cref{eq:x-clock} is the projection from a measured differential drift to the scalar velocity. It is conditional on the sensitivity difference $k_2-k_1$ \cite{Trivedi2025,Uzan2011}. The clock scale in Table~\ref{tab:local-bounds} corresponds to $|D|/H_0=O(10^{-8})$ at $95\%$ confidence. 

A coupling direction with $|\Delta k|\sim10^{-6}$ therefore gives the representative $|x|\lesssim10^{-2}$ regime, whereas a clock null direction with $\Delta k\simeq0$ gives no useful bound on $x$. The microscopic map below replaces the representative number by the direction dependent quantity $x_{\rm max}(\dvec)$. If $\phi$ is canonical and supplies the present dark energy density, then
\begin{equation}
        \rho_\phi=\frac12\dot\phi^2+V,
\end{equation}
\begin{equation}
        p_\phi=\frac12\dot\phi^2-V.
\end{equation}
These are the usual quintessence stress energy relations for a minimally normalized scalar dark energy component \cite{PeeblesRatra1988,CaldwellDaveSteinhardt1998}. We can write
\begin{equation}
        1+w_0=\frac{\rho_\phi+p_\phi}{\rho_\phi}=\frac{\dot\phi_0^2}{3\Mpl^2H_0^2\Ode}=\frac{x^2}{3\Ode}.
        \label{eq:canonical-w}
\end{equation}
For $x_{\rm max}\sim10^{-2}$ and $\Ode\simeq0.7$, this gives $1+w_0\lesssim 5\times10^{-5}.$ For a $P(X,\phi)$ scalar, with $X=\dot\phi^2/2$, the clock relation \cref{eq:x-clock} is unchanged, while
\begin{equation}
        1+w_0=\frac{2XP_X}{3\Mpl^2H_0^2\Ode} =\frac{P_X(X_0,\phi_0)x^2}{3\Ode}.
        \label{eq:noncanonical-w}
\end{equation}
The generalization follows the noncanonical or $k$-essence dark energy work, where $P_X$ gives the kinetic contribution to the stress tensor \cite{ArmendarizPiconMukhanovSteinhardt2000,ArmendarizPiconMukhanovSteinhardt2001,DeffayetPujolasSawickiVikman2010}. The conclusion of slow present motion therefore persists, although the translation from $x$ to $1+w_0$ acquires the factor $P_X$. For the universal conformal coupling in \cref{eq:einstein-action}, we define
\begin{equation}
        \beta(\phi)\equiv \Mpl\frac{\dd\ln A}{\dd\phi}.
\end{equation}
There is an equally important and physically equivalent way for matter to source the scalar. If the Einstein frame mass of a nonrelativistic species depends on $\phi$, then at fixed comoving particle number
\begin{equation}
 \rho_I(\phi)=m_I^E(\phi)n_I,
\end{equation}
\begin{equation}
     \frac{\partial\rho_I}{\partial\phi} = \frac{\partial\ln m_I^E}{\partial\phi}\rho_I .
\end{equation}
Thus a field dependent mass contributes to the scalar equation just as a conformal matter prefactor does. After absorbing the sign convention into $\beta_I$, the homogeneous equation is
\begin{equation}
        \ddot\phi+3H\dot\phi+V'(\phi) =\frac{1}{\Mpl}\sum_I\beta_I(\phi)\rho_I.
        \label{eq:eom-beta}
\end{equation}
Equivalently, over a local field interval on which the densities can be treated adiabatically, one may write
\begin{equation}
 V_{{\rm eff},\phi} = V_{,\phi} - \frac{1}{\Mpl}\sum_I\beta_I(\phi)\rho_I, \qquad \ddot\phi+3H\dot\phi+V_{{\rm eff},\phi}=0.
 \label{eq:effective-source-potential}
\end{equation}
This is the effective source potential referred to below. A scalar dependent prefactor multiplying a matter kinetic term and a scalar dependent mass should not be counted as two unrelated effects. 

Canonically normalizing the matter field moves part of the prefactor into masses and interaction vertices. The invariant macroscopic information is the scalar charge $\partial\ln m_A^E/\partial(\phi/\Mpl)$, which we construct in Sec.~\ref{sec:microscopic-map}. Field dependent dark matter masses are the standard language of coupled quintessence and interacting dark sector models \cite{Amendola2000}. The same observation has a useful connection to the distance conjecture. Along an asymptotic trajectory, a tower is expected to behave as
\begin{equation}
m_n(\phi) \sim m_n(\phi_{\rm ref}) \exp\left[ -a\,\frac{\Delta\phi}{\Mpl} \right], \qquad a=O(1), 
\label{eq:tower-mass}
\end{equation}
so each tower state carries an order one logarithmic scalar charge \cite{OoguriVafa2006,HeidenreichReeceRudelius2018}. If such states are populated and nonrelativistic, their densities enter \cref{eq:effective-source-potential}. Even when they are not populated, the tower becoming light lowers the EFT cutoff and eventually invalidates the finite field description. These are related but nonetheless different. 

The first is a local matter source, while the second is the distance conjecture's breakdown of the EFT. For the late universe the source sum may be split into visible baryons and cold dark matter. The universal matter limit is recovered by setting all $\beta_I$ equal. Since only the absolute allowed region for $\lambda_V$ will matter below, reversing the coupling sign convention amounts to $\beta_I\rightarrow-\beta_I$. The distance conjecture concerns the total controlled field space distance. The present local bound gives only a late time, one Hubble interval displacement, we have
\begin{equation}
        \Delta\phi_{H_0}\sim \frac{|\dot\phi_0|}{H_0}=|x|\Mpl.
\end{equation}
\begin{equation}
        \frac{\Delta\phi_{H_0}}{\Mpl}\lesssim x_{\rm max}\sim10^{-2}
\end{equation}
in the representative ultra slow regime. This local Hubble time excursion should not be mistaken for a complete distance conjecture test, which concerns parametrically large distances along the entire controlled EFT trajectory \cite{OoguriVafa2006,Palti2019}. Over an extended redshift interval, we have
\begin{equation}
\frac{\Delta\phi}{\Mpl}=\int_{a_i}^{1}x(a)\,\dd\ln a=\int_{0}^{z_i}\frac{x(z)}{1+z}\,\dd z .
\end{equation}
A scalar that is nearly frozen today may have moved substantially at earlier times. Reconstructing that history requires a potential, a coupling model and a cosmological background. With $N=\ln a$, the scalar equation becomes
\begin{equation}
\frac{\dd x}{\dd N}+\left(3+\frac{\dd\ln H}{\dd N}\right)x+\frac{V_{,\phi}}{H^2\Mpl}=3\sum_I\beta_I(\phi)\Omega_I .
\end{equation}
Together with $\dd\phi/\dd N=\Mpl x$ and the Friedmann equations, this system can be evolved backward from the locally constrained endpoint $(\phi_0,x_0)$. The measurements provide a present boundary condition, while
\begin{equation}
\frac{|\Delta\phi|}{\Mpl}=\left|\int_{N_i}^{0}x(N)\,\dd N\right|
\end{equation}
remains model dependent. Once the potential and couplings are specified, the local endpoint can nevertheless exclude otherwise possible histories. Evaluating \cref{eq:eom-beta} at $z\simeq0$, we define
\begin{equation}
        y\equiv \frac{\ddot\phi_0}{H_0^2\Mpl}.
\end{equation}
Since we have
\begin{equation}
        V'(\phi_0)=\frac{\beta_{b,0}\rho_{b,0}+\beta_{c,0}\rho_{c,0}}{\Mpl}-3H_0\dot\phi_0-\ddot\phi_0,
\end{equation}
using $\rho_{b,0}=3\Mpl^2H_0^2\Ob$, $\rho_{c,0}=3\Mpl^2H_0^2\Oc$ and $V(\phi_0)\simeq3\Mpl^2H_0^2\Ode$ we obtain
\begin{equation}
    \Mpl\frac{V'(\phi_0)}{V(\phi_0)}\simeq\frac{\Ob}{\Ode}\beta_{b,0}+\frac{\Oc}{\Ode}\beta_{c,0}-\frac{x}{\Ode}-\frac{y}{3\Ode}.
    \label{eq:Vprime-dimensionless}
\end{equation}
The single coupling expression with $\Om\beta_0$ is therefore the universal matter benchmark $\beta_{b,0}=\beta_{c,0}=\beta_0$. The visible sector interpretation of the local bounds instead sets the locally constrained term proportional to $\Ob\beta_{b,0}$ unless a separate dark matter coupling is specified.

The swampland gradient variable is
\begin{equation}
        \lambda_V\equiv \Mpl\frac{|V'(\phi_0)|}{V(\phi_0)}.
\end{equation}
In the slow acceleration limit $|y|\ll |x|$, \cref{eq:Vprime-dimensionless} reduces to
\begin{equation}
        \lambda_V\simeq\left|\frac{\Ob}{\Ode}\beta_{b,0}+\frac{\Oc}{\Ode}\beta_{c,0}-\frac{x}{\Ode}\right|
\end{equation}
with $|x|\leq x_{\rm max}$. In the visible sector projection $\beta_{c,0}=0$. Equivalently, for constant $\beta_{b,0}$, we have
\begin{equation}
        \max\left(0,\left|\frac{\Ob}{\Ode}\beta_{b,0}\right|-\frac{x_{\rm max}}{\Ode}\right)\leq \lambda_V\leq\left|\frac{\Ob}{\Ode}\beta_{b,0}\right|+\frac{x_{\rm max}}{\Ode}.
        \label{eq:lambdaV-band}
\end{equation}
For a sign convention with $\beta_{b,0}\geq0$ and away from the origin, this is
\begin{equation}
        \left|\lambda_V-\frac{\Ob}{\Ode}\beta_{b,0}\right|\leq \frac{x_{\rm max}}{\Ode}.
\end{equation}
The resulting band should be compared with cosmological swampland quintessence bounds which usually constrain $\lambda_V$ indirectly through expansion history observables rather than through local clock, LLR, PPN, and equivalence principle measurements \cite{llr1Williams:2004qba,llr2Williams:2004uw,llr3Merkowitz:2010kka,llr4Muller:2007zzb,llr5courde2017lunar,llr6alBurrage:2020jkj,llr7alMould:2014iga,RaveriHuSethi2018,AkramiKalloshLindeVardanyan2018}. 

In the local map, the visible sector contribution is therefore a baryon weighted term, while the dark matter contribution remains an independent direction unless a model for the dark sector coupling is imposed. If an acceleration prior $|y|\leq y_{\rm max}$ is retained, the right hand side of the band is replaced by
\begin{equation}
        \delta_{\lambda}=\frac{x_{\rm max}+y_{\rm max}/3}{\Ode}.
\end{equation}
It is worth emphasizing that the bound \(|y|\leq y_{\rm max}\) is not given by the local clock, LLR, PPN, or equivalence principle data. It is an additional dynamical prior. We leave \(y_{\rm max}\) unspecified. Every sharp gradient bound obtained in the slow acceleration limit should be understood as conditional on \(|y|\ll|x|\). 

Now, without a slow acceleration assumption, the gradient map broadens along the $y$ direction. This is the main nuisance direction in the local gradient inference, an unconstrained $y$ can imitate a larger local potential slope. For slow acceleration, the original de Sitter gradient criterion is immediate. The conjecture requires $\lambda_V\geq \cstar$. The local band can intersect this region only if
\begin{equation}
        \cstar\leq \left|\frac{\Ob}{\Ode}\beta_{b,0}\right|+\frac{x_{\rm max}}{\Ode}
\end{equation}
in the visible sector projection. If the visible matter coupling is bounded by $|\beta_{b,0}|\leq\beta_{\rm max}$, then a necessary condition for compatibility is
\begin{equation}
        \cstar\leq \frac{\Ob}{\Ode}\beta_{\rm max}+\frac{x_{\rm max}}{\Ode}.
\end{equation}
\begin{equation}
        \beta_{\rm max}\geq \frac{\Ode}{\Ob}\left(\cstar-\frac{x_{\rm max}}{\Ode}\right)
\end{equation}
is required if the gradient conjecture is to be satisfied through visible matter coupling without tuning the acceleration term. For $\Ob=0.049$, $\Oc=0.266$, $\Om=0.315$, $\Ode=0.685$, and $x_{\rm max}=10^{-2}$, we have ${x_{\rm max}}/{\Ode}=1.46\times10^{-2}$, and ${\Ode}/{\Ob}=13.98.$ Thus $\cstar=1$ requires $\beta_{\rm max}\gtrsim13.8$, while even $\cstar=0.1$ requires $\beta_{\rm max}\gtrsim1.19$. In the universal matter benchmark one instead uses $\Ode/\Om=2.17$, giving $\beta_{\rm max}\gtrsim2.14$ for $\cstar=1$ and $\beta_{\rm max}\gtrsim0.19$ for $\cstar=0.1$. 

Such large unscreened visible sector couplings are not compatible with the post-Newtonian and equivalence principle interpretation of a long-range universally coupled scalar. Large unscreened scalar couplings usually require screening, a hidden sector coupling, or another way of evading local bounds \cite{BertottiIessTortora2003,WagnerSchlammingerGundlachAdelberger2012,Touboul2022,KhouryWeltman2004,JoyceJainKhouryTrodden2015}. In the small coupling regime $|\beta_{b,0}|\ll1$ and $|\beta_{c,0}|\ll1$ we obtain
\begin{equation}
        \lambda_V\lesssim \frac{x_{\rm max}}{\Ode}=O(10^{-2}),
\end{equation}
which is a local version of the well known tension between $w\simeq-1$, small scalar motion, and an order one de Sitter gradient coefficient in quintessence models \cite{AgrawalObiedSteinhardtVafa2018,RaveriHuSethi2018}. The refined de Sitter alternative introduces
\begin{equation}
        \eta_V\equiv \Mpl^2\frac{V''(\phi_0)}{V(\phi_0)}.
\end{equation}
The variable $\eta_V$ is the single field version of the smallest Hessian eigenvalue appearing in the refined de Sitter conjecture \cite{GargKrishnan2018,OoguriPaltiShiuVafa2018}. 

If the gradient condition fails because $\lambda_V\ll1$, the single field hilltop realization can still satisfy the refined condition through a tachyonic Hessian direction. In the growing mode branch, the local velocity bound requires $|\phi_0|/\Mpl\lesssim x_{\rm max}/\lambda_+(|\eta_V|)$. Thus, for $|\eta_V|=O(1)$, the scalar must currently remain very close to the hilltop, unless the growing mode is suppressed through a tuning of the initial conditions. The detailed evaluation of the gradient variable at the hilltop is given in Appendix~\ref{app:hilltop-evaluation}.

\section{Coupling consistent microscopic map}\label{sec:microscopic-map}

The previous section used the local quantities $x$, $\beta_{b,0}$, $\beta_{c,0}$, $F(\phi)$, $\alpha_{M,0}$, and the clock coefficients $k_i$ as phenomenological inputs. A more restrictive formulation is obtained by requiring the visible sector quantities to descend from one microscopic coupling vector while retaining a possible dark matter coupling as a separate source direction. We now construct this map. 

This removes the ambiguity that would otherwise allow the clock sensitivities, the matter coupling, the equivalence principle charges, and the Planck mass running to be chosen independently. Let
\begin{equation}
 \varphi \equiv \frac{\phi-\phi_0}{\Mpl}
\end{equation}
be the dimensionless local displacement of the normalized Einstein frame scalar. In the Jordan frame matter description, the Standard Model parameters $q_a(\phi)$ depend on this same scalar coordinate after the normalization convention of Sec.~\ref{sec:local-constraints} has been imposed. Around the present value $\phi_0$, we define $ d_F \equiv
({\partial F/\partial\varphi})/{F}|_0,$ $ d_e \equiv ({\partial \alpha/\partial\varphi})/{\alpha}|_0,$ $d_g \equiv ({\partial \Lambda_{\rm QCD}/\partial\varphi})/
 {\Lambda_{\rm QCD}}|_0$ and $ d_{m_i} \equiv ({\partial m_i/\partial\varphi})/{m_i}|_0.$ Equivalently, at linear order, it can be said that    
\begin{equation}
 \mathcal L_{\rm int}^{J} = \varphi
 \left[ \frac{d_e}{4e^2}F_{\mu\nu}F^{\mu\nu} - d_g\frac{\beta_3}{2g_3}G^a_{\mu\nu}G_a^{\mu\nu} - \sum_{i=e,u,d} d_{m_i}m_i\bar\psi_i\psi_i + \cdots \right]
 \label{exp}
\end{equation}
up to sign conventions for the gauge kinetic terms. This is the light dilaton coupling basis of Damour-Donoghue \cite{DamourPolyakov1994, DamourDonoghuePhen,DamourDonoghueEP}. It is useful to replace $m_u$ and $m_d$ by
\begin{equation}
        \hat m\equiv\frac{m_u+m_d}{2},
\end{equation}
\begin{equation}
        \delta m\equiv m_d-m_u,
\end{equation}
and to define the QCD subtracted couplings
\begin{equation}
 \bar d_{m_e}\equiv d_{m_e}-d_g,
\end{equation}
\begin{equation}
 \bar d_{\hat m}\equiv d_{\hat m}-d_g,
\end{equation}
\begin{equation}
 \bar d_{\delta m}\equiv d_{\delta m}-d_g .
\end{equation}
The subtraction is important because a common rescaling of all masses by $\Lambda_{\rm QCD}$ is largely universal, whereas equivalence principle violation is controlled by composition dependent deviations from this universal scaling. For a macroscopic body $A$, we define its Jordan frame scalar charge by $q_A^J \equiv {\partial \ln m_A^J}/{\partial \varphi}|_0 .$ To leading order in the nuclear binding expansion,
\begin{equation}
 q_A^J = d_g +\bar d_{\hat m} Q_{\hat m}^A +\bar d_{\delta m} Q_{\delta m}^A +\bar d_{m_e} Q_{m_e}^A +d_e Q_e^A ,
\end{equation}
where the $Q_i^A$ are nuclear and electromagnetic dilaton charges. Their semiempirical forms are tabulated in \cite{DamourDonoghuePhen,DamourDonoghueEP} and a complete analysis should use the isotope dependent charges rather than treating $Q_i^A$ as adjustable parameters. Passing to the Einstein frame with $g_{\mu\nu}^{E}=F(\phi)\tilde g_{\mu\nu}$, the Einstein frame mass is $ m_A^{E}(\phi)=F^{-1/2}(\phi)m_A^J(\phi).$ Hence the scalar charge sourcing the fifth force is
\begin{equation}
\alpha_A \equiv \left. \frac{\partial \ln m_A^{E}}{\partial \varphi} \right|_0 = -\frac{1}{2}d_F +d_g +\bar d_{\hat m} Q_{\hat m}^A +\bar d_{\delta m} Q_{\delta m}^A +\bar d_{m_e} Q_{m_e}^A +d_e Q_e^A .
\end{equation}
The universal Planck mass coupling $d_F$, the QCD coupling $d_g$ and the composition dependent charges therefore enter the same macroscopic scalar charge. For the baryonic material relevant to local tests, we define the visible matter weighted coupling
\begin{equation}
 \beta_{\rm m} \equiv \sum_A f_A \alpha_A .
\end{equation}
Here the sum is over the visible nonrelativistic mixture used in the local source or test body model. A dark matter coupling is a separate cosmological parameter, denoted $\beta_c$, unless one assumes the universal matter limit. In the notation of the previous section, $\beta_{b,0}$ should therefore be identified with $\beta_{\rm m}$. It is not an independent number once the microscopic vector
\begin{equation}
        \dvec=(d_F,d_g,d_e,d_{m_e},d_{\hat m},d_{\delta m},\ldots)
        \label{vector}
\end{equation}
has been specified. The EFT of dark energy Planck mass running is also determined by the same local expansion
\begin{equation}
 \alpha_M \equiv \frac{\dd\ln M_*^2}{\dd\ln a} = \frac{\dd\ln F}{\dd\ln a} = d_F x,
\end{equation}
thus $F(\phi)$, $x$, and $\alpha_M$ cannot be varied independently. At linear order, $\alpha_{M,0}=d_Fx.$ For atomic clocks, a transition $I$ has the local expansion
\begin{equation}
 \frac{\partial \ln \nu_I}{\partial \varphi} \equiv k_I(\dvec) = K_\alpha^I d_e +K_e^I(d_{m_e}-d_g) +K_{\hat m}^I(d_{\hat m}-d_g) +K_{\delta m}^I(d_{\delta m}-d_g) +\cdots
\end{equation}
where the $K_a^I$ are atomic structure sensitivity coefficients. For a ratio of two clocks, we have
\begin{equation}
 D_{IJ} \equiv \frac{\dd}{\dd t}\ln\left(\frac{\nu_I}{\nu_J}\right) = H_0 x\,\Delta k_{IJ}(\dvec),
\end{equation}
\begin{equation}
 \Delta k_{IJ}(\dvec)\equiv k_I(\dvec)-k_J(\dvec) .
\end{equation}
Therefore the clock bound is a bound on the product $x\Delta k_{IJ}(\dvec)$. For a measured mean $D_{IJ}^{\rm obs}$ with Gaussian standard uncertainty $\sigma_{IJ}$, define the conservative absolute envelope
\begin{equation}
 D_{IJ}^{\rm lim}({\rm CL})
 \equiv
 |D_{IJ}^{\rm obs}|+z_{\rm CL}\sigma_{IJ},
\end{equation}
where $z_{\rm CL}=1$ gives the central $68.3\%$ interval and $z_{\rm CL}=1.96$ gives the central $95\%$ interval. Then
\begin{equation}
 |x| \le x_{\rm clk}(\dvec) \equiv \min_{IJ} \frac{D_{IJ}^{\rm lim}({\rm CL})}{H_0|\Delta k_{IJ}(\dvec)|},
      \label{xbound}
\end{equation}
Stating the confidence level through \(z_{\rm CL}\) makes the numerical prescription unambiguous. \Cref{xbound} is a convenient conservative envelope and the Gaussian likelihood used later is preferable when central values, correlations and non-Gaussian errors are retained. If a coupling direction satisfies $\Delta k_{IJ}(\dvec)\simeq0$, that clock pair loses sensitivity to scalar motion along that direction. 

For equivalence principle tests, the scalar mediated force between a source $S$ and a test body $A$ is proportional to $G_{AS}=G_*F_0^{-1}(1+\alpha_A\alpha_S)$. The E\"otv\"os ratio for two test bodies $A,B$ falling toward the same source $S$ is therefore
\begin{equation}
 \eta_{AB}^{(S)} \equiv 2\frac{a_A-a_B}{a_A+a_B} \simeq (\alpha_A-\alpha_B)\alpha_S,
\end{equation}
up to higher orders in the small charges. Equivalence principle searches therefore impose
\begin{equation}
 \left| \left[ \bar d_{\hat m}\Delta Q_{\hat m}^{AB} +\bar d_{\delta m}\Delta Q_{\delta m}^{AB} +\bar d_{m_e}\Delta Q_{m_e}^{AB} +d_e\Delta Q_e^{AB} \right]\alpha_S(\dvec) \right| \le \eta_{AB,{\rm max}}^{(S)} .
 \label{eqbound}
\end{equation}
The universal piece $-d_F/2+d_g$ cancels from $\alpha_A-\alpha_B$, but it remains in the source charge $\alpha_S$. The same microscopic couplings that determine clock drifts therefore also determine the composition dependence of free fall. 

We can now connect the source charge to the PPN metric introduced in \cref{eq:ppn-metric}. In the effectively massless limit, and for a source whose composition is fixed, matching the scalar exchange correction to the weak field metric gives
\begin{equation}
 \gamma_{\rm PPN}^{(S)}-1 \simeq -\frac{2\alpha_S^2}{1+\alpha_S^2}.
 \label{eq:gamma-source-charge}
\end{equation}
For a finite range scalar this relation acquires the usual distance dependent Yukawa suppression, so the Cassini number cannot be applied without checking that the scalar range is at least of Solar System size. In the long range case, the Sun supplies the relevant source charge and the measured bound implies
\begin{equation}
 \frac{2\alpha_\odot^2}{1+\alpha_\odot^2} \le |\gamma_{\rm PPN}-1|_{\rm max}.
\end{equation}
For $\alpha_\odot^2\ll1$, we approximate
\begin{equation}
 |\alpha_\odot| \lesssim \left(\frac{|\gamma_{\rm PPN}-1|_{\rm max}}{2}\right)^{1/2}.
 \label{eq:alpha}
\end{equation}
We also have
\begin{equation}
 \frac{\dot G_{AB}}{G_{AB}} = -H_0\alpha_{M,0} +H_0x \frac{ \alpha_A'\alpha_B+\alpha_A\alpha_B'}{1+\alpha_A\alpha_B}
\end{equation}
where a prime denotes $\dd/\dd\varphi$. For the linear coupling basis used above, $\alpha_A'$ is second order in the local EFT expansion and may be neglected unless nonlinear dilaton couplings are retained. Then
\begin{equation}
 \frac{\dot G_{AB}}{G_{AB}} \simeq -H_0 d_F x = -H_0\alpha_{M,0}.
\end{equation}
Hence the LLR prior is not independently a prior on $x$. It is a prior on $d_Fx$. In the numerical examples we use
\begin{equation}
 |d_Fx|\leq\epsilon_G,
 \qquad
 \epsilon_G\equiv\epsilon_G^{95}\simeq3.5\times10^{-4},
\end{equation}
as obtained in \cref{eq:epsilonG95}.
The dark energy equation of state remains as in \cref{eq:canonical-w}
for a canonical scalar. For a noncanonical $P(X,\phi)$ theory this is multiplied by $P_X(X_0,\phi_0)$ as in \cref{eq:noncanonical-w}. The scalar equation of motion in the Einstein frame is
\begin{equation}
 \ddot\phi+3H\dot\phi+V'(\phi) = \frac{\beta_{\rm m}(\dvec)\rho_b+\beta_c\rho_c}{\Mpl}
\end{equation}
with the same sign convention as \cref{eq:eom-beta}. Evaluating at $z=0$,
\begin{equation} 
\frac{\Mpl V'_0}{V_0} = \frac{\Ob}{\Ode}\beta_{\rm m}(\dvec) +\frac{\Oc}{\Ode}\beta_c -\frac{x}{\Ode} -\frac{y}{3\Ode} . 
\end{equation} 
\begin{equation} \lambda_V(\dvec,\beta_c,x,y) = \left| \frac{\Ob}{\Ode}\beta_{\rm m}(\dvec) +\frac{\Oc}{\Ode}\beta_c -\frac{x}{\Ode} -\frac{y}{3\Ode} \right| . 
\label{eq:lambdaV-micro} 
\end{equation}
In the universal matter coupling \(\beta_c=\beta_{\rm m}\equiv\beta\), the source in \cref{eq:lambdaV-micro} becomes \((\Omega_{m,0}/\Omega_{DE,0})\beta\simeq0.46\beta\), rather than the baryon weighted contribution \((\Omega_{b,0}/\Omega_{DE,0})\beta_{\rm m}\simeq0.0715\beta_{\rm m}\). Universality therefore enhances the matter contribution to \(\lambda_V\) by a factor of approximately \(6.4\). The gain is real, but it is not enough. The same universal coupling controls the scalar charge of Solar System bodies and is directly constrained by post-Newtonian and fifth force tests. For an unscreened long range scalar, Cassini requires \(|\beta|\) to be at most a few \(\times10^{-3}\), leaving a universal matter contribution of only order \(10^{-3}\). Neglecting \(x\) and \(y\), satisfying \(\cstar=1\) would instead require \(|\beta|\gtrsim\Omega_{DE,0}/\Omega_{m,0}\simeq2.17\), while even \(\cstar=0.1\) requires \(|\beta|\gtrsim0.2\). Both are incompatible with the unscreened universal limit.

Universality thus reduces the cosmological density weighting penalty but does not remove the tension with the de Sitter slope conjecture. Compared with a model in which \(\beta_c\) is an independent dark sector coupling, it is also less flexible: the dark matter source cannot be increased without increasing the locally visible charge. The visible matter coupling $\beta_{\rm m}$, clock coefficients, E\"otv\"os charges and PPN source charge are all projections of $\dvec$, whereas $\beta_c$ remains independent unless universality is imposed. Combining the clock and LLR bounds gives the direction dependent velocity prior
\begin{equation}
|x| \le x_{\rm max}(\dvec) \equiv \min\left[ x_{\rm clk}(\dvec), \frac{\epsilon_G}{|d_F|} \right],
\end{equation}
To illustrate the directional dependence, consider the truncated coupling basis
\begin{equation}
\textbf{d}=(d_F,d_g,d_e,d_{m_e},d_{\hat m},d_{\delta m}).    
\end{equation}
An electromagnetic direction such as \(\mathbf d_{\rm em}=(0,0,10^{-6},0,0,0)\) gives \(\Delta k_{IJ}\simeq10^{-6}\Delta K_{\alpha}^{IJ}\). For a clock pair with \(|\Delta K_{\alpha}^{IJ}|\sim1\) and \(D_{IJ}^{\rm lim}(95\%)/H_0\sim10^{-8}\), \cref{xbound} gives \(x_{\rm clk}\sim10^{-2}\). Increasing the same coupling to \(d_e=10^{-4}\) tightens the clock limit to \(x_{\rm clk}\sim10^{-4}\), provided that the direction also satisfies the equivalence principle constraint in \cref{eqbound}.

By contrast, consider the approximately universal rescaling direction
\begin{equation}
\mathbf d_{\rm univ}
=(0,10^{-3},0,10^{-3},10^{-3},10^{-3}).
\end{equation}
It has \(\bar d_{m_e}=\bar d_{\hat m}=\bar d_{\delta m}=0\), and hence \(\Delta k_{IJ}\simeq0\) and \(\alpha_A-\alpha_B\simeq0\) at leading order. Since \(d_F=0\), LLR places no direct bound on \(x\), although the residual universal charge \(\alpha_A\simeq10^{-3}\) remains subject to PPN constraints. The pure Planck mass direction \(\mathbf d_F=(10^{-3},0,0,0,0,0)\) is also invisible to clock ratios. With the current \(\epsilon_G\simeq3.5\times10^{-4}\), LLR alone gives \(|x|\lesssim0.35\) along this direction, so cosmological energy density and equation of state information can dominate. 

These examples show why \(|x|\lesssim10^{-2}\) is representative rather than universal, it applies to coupling vectors with an appreciable projection onto the clock sensitivity matrix, while clock null directions with small \(d_F\) can leave the present velocity much less constrained. Independently of the velocity projection, the coupling vector must satisfy
\begin{equation}
 |\eta_{AB}^{(S)}(\dvec)|\le \eta_{AB,{\rm max}}^{(S)},
\end{equation}
\begin{equation}
 |\gamma_{\rm PPN}^{(\odot)}(\dvec)-1|
 \le
 |\gamma_{\rm PPN}-1|_{\rm max}.
 \label{PPNBound}
\end{equation}
Note also that the microscopic basis in \crefrange{exp}{vector} is not incompatible with screening. It parametrizes the couplings of the scalar to Standard Model operators, whereas screening is determined by the nonlinear scalar dynamics, its potential, and the environmental dependence of the background field. In chameleon, symmetron, and environment dependent dilaton realizations, the same microscopic coupling directions may be present, but the field profile, effective mass, local coupling, or macroscopic scalar charge can become density dependent. The observable combinations \(x_{\rm loc}\Delta k_{IJ}(\mathbf d)\), \(\alpha_A^{\rm eff}\alpha_S^{\rm eff}\) and \((d_Fx)_{\rm loc}\) entering \crefrange{xbound}{PPNBound} must then be evaluated using the screened local solution.

Consequently, the unscreened constraints derived here cannot be applied directly to the bare coupling vector \(\mathbf d\). Vainshtein or kinetic screening is different because derivative nonlinearities primarily suppress spatially sourced fifth forces and need not suppress the cosmological time dependence entering clock drifts or \(\dot G/G\) and whether those observables are screened is therefore model dependent \cite{sc1burrage2014screening, sc2fischer2024screened, sc3shibata2023properties}. Our results should accordingly not be read as excluding every unscreened light scalar.

They exclude, or strongly restrict, the more specific possibility that a locally visible, unscreened scalar obtains an order one de Sitter gradient predominantly through visible sector matter couplings. If that is imposed, screening or another mechanism that suppresses the local response becomes necessary. Now the allowed region is the projection of a constrained region in microscopic coupling space. For an acceleration prior $|y|\leq y_{\rm max}$, the allowed band in the $(\lambda_V,\beta_{\rm m})$ plane at a constant $\beta_c$ is
\begin{equation} 
\left| \lambda_V - \frac{\Ob}{\Ode}\beta_{\rm m} - \frac{\Oc}{\Ode}\beta_c \right| \le \frac{x_{\rm max}(\dvec)+y_{\rm max}/3} {\Ode} . 
\end{equation} 
Equivalently, for constant $\beta_{\rm m}$ and $\beta_c$, 
\begin{equation} \lambda_{V,\rm min} = \max\left[ 0, \left| \frac{\Ob}{\Ode}\beta_{\rm m} + \frac{\Oc}{\Ode}\beta_c \right| - \frac{x_{\rm max}(\dvec)+y_{\rm max}/3} {\Ode} \right], 
\end{equation} 
\begin{equation} \lambda_{V,\rm max} = \left| \frac{\Ob}{\Ode}\beta_{\rm m} + \frac{\Oc}{\Ode}\beta_c \right| + \frac{x_{\rm max}(\dvec)+y_{\rm max}/3} {\Ode} . 
\end{equation} The contour in the $(\lambda_V,y)$ plane is similarly 
\begin{equation} \lambda_{V,\rm min}(y) = \max\left[ 0, \left| \frac{\Ob}{\Ode}\beta_{\rm m}(\dvec) + \frac{\Oc}{\Ode}\beta_c - \frac{y}{3\Ode} \right| - \frac{x_{\rm max}(\dvec)}{\Ode} \right], 
\label{eq:lambdaV-y-min} 
\end{equation} 
\begin{equation} \lambda_{V,\rm max}(y) = \left| \frac{\Ob}{\Ode}\beta_{\rm m}(\dvec) + \frac{\Oc}{\Ode}\beta_c - \frac{y}{3\Ode} \right| + \frac{x_{\rm max}(\dvec)}{\Ode} . 
\label{eq:lambdaV-y-max} 
\end{equation}
\Cref{eq:lambdaV-y-min,eq:lambdaV-y-max} show how an unconstrained acceleration direction broadens the gradient conjecture map. The earlier representative band is recovered only after replacing $x_{\rm max}(\dvec)$ by a constant number and choosing $y_{\rm max}$. The local observables above do not determine $\eta_V$ by themselves. They constrain it only after specifying the local form of the potential. Near a hilltop at $\phi=\phi_h$ for $\eta_V<0$,
\begin{equation}
 V(\phi)= V_0 \left[1-\frac{|\eta_V|}{2} \left(\frac{\phi-\phi_h}{\Mpl}\right)^2 +\cdots \right].
 \label{hilltop}
\end{equation}
The linearized local equation gives the form derived in Appendix~\ref{app:hilltop-evaluation}, as in \cref{eq:lambda-pm}, so that, in the growing mode branch
\begin{equation}
|x| \simeq \lambda_+(|\eta_V|) \left| \frac{\phi_0-\phi_h}{\Mpl} \right| .
\end{equation}
The coupling consistent hilltop contour is therefore
\begin{equation}
 \left| \frac{\phi_0-\phi_h}{\Mpl} \right| \le \frac{x_{\rm max}(\dvec)}{\lambda_+(|\eta_V|)} .
\end{equation}
\begin{figure}[t]
\centering
\includegraphics[width=0.92\linewidth]{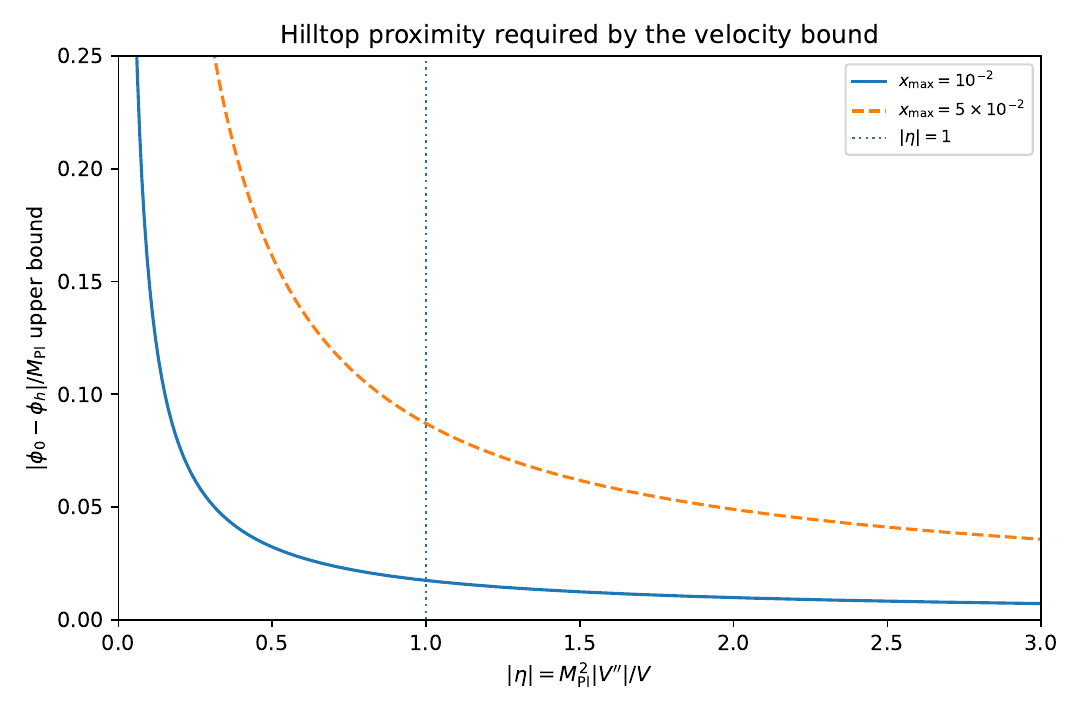}
\caption{Hilltop proximity bound obtained from the growing mode relation $|x|\simeq\lambda_+(|\eta_V|)|\phi_0-\phi_h|/\Mpl$. Curves show $|\phi_0-\phi_h|/\Mpl\leq x_{\rm max}/\lambda_+(|\eta_V|)$ for representative velocity bounds. For $|\eta_V|=O(1)$, the allowed displacement is of order $10^{-2}$-$10^{-1}$ for the benchmark values used in the text.}
\label{fig:hilltop-proximity}
\end{figure}Figure~\ref{fig:hilltop-proximity} displays this proximity requirement for representative values of $x_{\rm max}$. The divergence at small $|\eta_V|$ reflects the fact that a shallow tachyonic direction produces little present motion, while an order one refined de Sitter curvature requires the field to sit close to the hilltop if the local velocity is small. Since $x_{\rm max}$ itself depends on the microscopic coupling direction, a refined de Sitter compatible hilltop is allowed only along coupling directions that are consistent with clocks, LLR, PPN tests, and equivalence principle tests. This result should not be interpreted as a model independent experimental test, much less a falsification of the refined de Sitter conjecture. 

The conjecture requires an \(\mathcal{O}(1)\) tachyonic Hessian direction when the gradient is small but it does not by itself require the cosmological field to occupy an arbitrarily small neighborhood of the maximum at the present epoch, nor does it constrain the relative amplitudes of the growing and decaying modes. Local bounds give this additional dynamical information. Once one assumes an unscreened single field hilltop realization and a generic growing mode evolution, \(|x|\simeq\lambda_{+}(|\eta_V|)|\phi_0-\phi_h|/M_{\rm Pl}\) so the clock bound becomes the condition that \(|\phi_0-\phi_h|/M_{\rm Pl}\lesssim x_{\max}(\mathbf d)/\lambda_{+}(|\eta_V|)\). For \(|\eta_V|=1\) and \(x_{\max}=10^{-2}\), this requires a displacement below approximately \(1.7\times10^{-2}M_{\rm Pl}\). Part of this proximity is therefore intrinsic to obtaining a small gradient near a hilltop. 

However, the local velocity constraint makes the requirement sharper and shows that generic initial conditions may be incompatible with the locally allowed motion unless the field is sufficiently close to the maximum, the growing mode is tuned to be small, or the assumptions of an unscreened single field description are relaxed. We therefore regard the result primarily as a conditional consistency and naturalness constraint on refined de Sitter realizations. For a future material dependent likelihood one may use the form
\begin{equation}
\begin{aligned}
\chi^2_{\rm loc}(\dvec,x) =& \sum_{IJ} \frac{ \left[ D_{IJ}^{\rm obs} - H_0 x \Delta k_{IJ}(\dvec) \right]^2 }{\sigma_{IJ}^2} + \sum_{AB,S} \frac{ \left[ \eta_{AB}^{(S),{\rm obs}} - (\alpha_A-\alpha_B)\alpha_S \right]^2 }{\sigma_{AB,S}^2} \\ &+ \frac{ \left[ (\gamma_{\rm PPN}-1)_{\rm obs} + \frac{2\alpha_\odot^2}{1+\alpha_\odot^2} \right]^2 }{\sigma_\gamma^2} + \frac{ \left[ (\dot G/G)_{\rm obs} + H_0 d_F x \right]^2 }{\sigma_G^2} \,.
\label{chi}
\end{aligned}
\end{equation}
The corresponding region in swampland variables is obtained after adjoining explicit priors or dynamical and cosmological likelihoods for $\beta_c$, $y$, and $\eta_V$, and then mapping each combined point through
\begin{equation}
(\dvec,\beta_c,x,y,\eta_V) \to \left( \lambda_V(\dvec,\beta_c,x,y), \eta_V, \beta_{\rm m}(\dvec), \beta_c, x, y, \alpha_{M,0}=d_Fx \right).
 \label{map}
\end{equation}
We note here that a numerical implementation of \cref{chi} is beyond the scope of the present study. A future analysis would combine measured clock drifts and their atomic sensitivity coefficients, material dependent dilaton charges for equivalence principle experiments, the Solar System source composition entering the PPN constraint, and the LLR bound on \(\dot G/G\), including the corresponding experimental covariances where available. The local likelihood should first be sampled in the underlying parameter space \((\mathbf d,x)\) since \(\beta_m\), \(\alpha_M\), the clock sensitivities, and the macroscopic scalar charges are correlated projections of the same microscopic coupling vector rather than independent phenomenological parameters. 

It may then be combined with priors or cosmological and dynamical likelihoods for \(\beta_c\), \(y\), and the potential parameters, because the local likelihood is otherwise flat along those directions. Each point in the combined parameter space can then be mapped through \cref{map} to obtain marginalized allowed regions in \((\lambda_V,\beta_m)\) and related parameter planes. A constraint in the complete \((\lambda_V,\eta_V,\beta_m)\) space requires an additional specification of the local potential because the local observables entering \cref{chi} do not constrain \(\eta_V\) directly. For example, imposing the hilltop expansion in \cref{hilltop} and evolving the scalar equation would relate \(\eta_V\) to the present displacement and velocity, permitting a model dependent likelihood in \((\lambda_V,\eta_V,\beta_m)\). 

We leave this material dependent numerical analysis to future work and the purpose of the present paper is to establish the coupling consistent map and identify the assumptions and degeneracy directions that such an analysis must retain.

\section{Representative local scales}

The previous section gives the local map without requiring a separate numerical analysis. Its physical interpretation is already visible from the relative size of the terms in \cref{eq:lambdaV-micro}. Using representative late time values from the Planck 2018 base $\Lambda$CDM analysis \cite{Planck2018VI},
\begin{equation}
        \Ob=0.049,
\end{equation}
\begin{equation}
        \Oc=0.266,
\end{equation}
\begin{equation}
        \Om=\Ob+\Oc=0.315,
\end{equation}
\begin{equation}
        \Ode=0.685,
\end{equation}
we have
\begin{equation}
        \frac{\Ob}{\Ode}=0.0715,
\end{equation}
\begin{equation}
        \frac{\Oc}{\Ode}=0.3883,
\end{equation}
\begin{equation}
        \frac{\Om}{\Ode}=0.4599.
\end{equation}
Thus the visible sector part of the homogeneous source is numerically suppressed relative to the total matter benchmark. This distinction is important because local clock, PPN, LLR, and equivalence principle probes constrain baryonic scalar charges directly.

A cold dark matter coupling enters through the separate term $(\Oc/\Ode)\beta_c$ and is not fixed by those visible sector measurements. For an approximately universal visible scalar charge, the Cassini bound on $\gamma_{\rm PPN}-1$ implies \cref{eq:alpha} which is a few $\times10^{-3}$ for the usual Solar System bound. Identifying the corresponding visible matter coupling with this scale gives
\begin{equation}
        \frac{\Ob}{\Ode}|\beta_{\rm m}|
        \lesssim
        {\rm few}\times10^{-4}.
\end{equation}
Composition dependent visible sector directions are more strongly restricted by equivalence principle searches, so their baryon weighted contribution is smaller still. These estimates are not a substitute for the material dependent likelihood described above. They simply show, before any elaborate fit, why the visible sector cannot by itself raise $\lambda_V$ to order unity in the unscreened local theory.

In the conservative visible sector projection $\beta_c=0$, \cref{eq:lambdaV-micro} shows that the visible coupling term is normally subdominant to the velocity and acceleration terms. If $|x|\lesssim10^{-2}$ and $|y|\ll|x|$,
\begin{equation}
 \frac{|x|}{\Ode}\lesssim1.5\times10^{-2}.
\end{equation}
The possible escape routes are therefore clear, that a significant acceleration term $y$, a dark sector coupling $\beta_c$, screening, a locally invisible coupling direction, or a departure from the single field unscreened assumptions. Finally, Table~\ref{tab:local-bounds} constrains the first time derivative of $F$, not its second derivative. From \cref{eq:alphaM-def}, we define
\begin{equation}
        A_F\equiv \frac{\dot F}{F}=H\alpha_M.
\end{equation}
\begin{equation}
        \dot\alpha_M
        =\frac{1}{H}\left(\frac{\ddot F}{F}-\frac{\dot F^2}{F^2}\right)
        -\alpha_M\frac{\dot H}{H}.
\end{equation}
\begin{equation}
        \frac{\dot\alpha_M}{H}
        =\frac{\ddot F}{H^2F}-\alpha_M^2-\alpha_M\frac{\dot H}{H^2}.
\end{equation}
A bound on $\dot G/G$ constrains $\dot F/F$, hence $\alpha_M$. It does not, by itself, determine $\ddot F/F$. Therefore $\dot\alpha_M$ remains a degenerate direction unless an independent observable sensitive to $\ddot G/G$ is included or a model for $F(\phi)$ is imposed. For example, expanding locally in the dimensionless $F$ convention
\begin{equation}
        F(\phi)=1+f_1\frac{\phi-\phi_0}{\Mpl}
        +\frac12 f_2\left(\frac{\phi-\phi_0}{\Mpl}\right)^2+\cdots.
\end{equation}
At the present point, $\alpha_{M,0}=f_1x$, and
\begin{equation}
        \frac{\dot\alpha_{M,0}}{H_0}
        \simeq f_1y+f_2x^2-f_1^2x^2-\alpha_{M,0}\frac{\dot H_0}{H_0^2}.
\end{equation}
Since $x^2\lesssim10^{-4}$ in the representative ultra slow regime, the $f_2x^2$ term is small unless $f_2$ is large, but the $f_1y$ term is not fixed without an acceleration prior. Hence the local inputs produce a constrained sheet in $(\alpha_M,x,\beta_b,\beta_c,\lambda_V,\eta_V,\phi)$ with a model dependent extension along the $\dot\alpha_M$ direction.

\section{Conclusion}

We have recast the Solar System and atomic clock problem as a local map from experimental observables to swampland variables. Local measurements do not constrain the scalar velocity alone. They constrain $x$ multiplied by microscopic coupling directions, together with source dependent and composition dependent scalar charges. With the present LLR input, $|d_Fx|=|\alpha_{M,0}|\lesssim3.5\times10^{-4}$ at the conservative $95\%$ level used here. Clock visible directions can give the representative $|x|\lesssim10^{-2}$ regime, but clock null directions need not. The statement is
\begin{equation}
 |x| \leq x_{\rm max}(\dvec) = \min\left[ x_{\rm clk}(\dvec), \frac{\epsilon_G}{|d_F|} \right],
\end{equation}
supplemented by PPN and equivalence principle constraints on $\dvec$. For a canonical scalar supplying the present dark energy density, $1+w_0=x^2/(3\Ode)$, so the representative clock visible regime gives $1+w_0\lesssim10^{-4}$--$10^{-5}$. 

The one Hubble time displacement is similarly small, $\Delta\phi_{H_0}/\Mpl\sim|x|$, but this is not a global distance conjecture test. A field frozen today may have moved earlier, and a distance conjecture tower introduces two effects, populated field dependent masses source the local scalar equation, while the tower becoming light eventually lowers the EFT cutoff. The de Sitter statement is most transparent in the notation used here. The potential diagnostic $\lambda_V$ is compared with the conjectural constant $\cstar$, they are not the same object. At $z=0$, \cref{eq:lambdaV-micro} shows that the visible source is weighted by $\Ob/\Ode\simeq0.0715$. 

Even a nearly universal visible charge at the PPN limit contributes only a few $\times10^{-4}$ to $\lambda_V$, and composition dependent directions are more tightly constrained by MICROSCOPE. In the slow acceleration, small visible coupling and $\beta_c=0$ limit, the representative result is therefore $\lambda_V\lesssim O(10^{-2})$. Reaching $\lambda_V\geq\cstar$ requires physics not fixed by the visible local bounds: a significant $y$, a dark or hidden sector coupling, screening, a locally invisible direction, tuned dynamics, a smaller conjectural coefficient, or a departure from the single field setup. 

The refined alternative remains possible, but it is also conditional. In the single field growing mode hilltop realization,
\begin{equation}
\frac{|\phi_0-\phi_h|}{\Mpl} \lesssim \frac{x_{\rm max}(\dvec)}{\lambda_+(|\eta_V|)}, \qquad \lambda_+(|\eta_V|) =
 \frac{-3+\sqrt{9+12\Ode|\eta_V|}}{2}.
\end{equation}
For $|\eta_V|=1$ and $x_{\rm max}=10^{-2}$, the displacement must be below approximately $1.7\times10^{-2}\Mpl$. This is a consistency and naturalness statement, not a model independent falsification of the refined conjecture. A locally visible, unscreened dark energy scalar cannot obtain an order one de Sitter slope naturally from visible sector matter coupling alone.

\appendix
\section{Evaluation of the gradient variable at the hilltop}\label{app:hilltop-evaluation}

If the gradient condition fails because $\lambda_V\ll1$, the refined conjecture attempts to retain compatibility by requiring $\eta_V\lesssim -\etastar$. The local velocity bound makes this difficult. To quantify the tuning, we expand the potential near a local maximum at $\phi=0$, that is
\begin{equation}
        V(\phi)=V_0-\frac12 |m^2|\phi^2+\cdots,
\end{equation}
where  $V_0\simeq3\Mpl^2H_0^2\Ode$. Hilltop and unstable de Sitter examples are the cases for which the refined conjecture was introduced since a small slope can coexist with a tachyonic Hessian direction \cite{GargKrishnan2018,OoguriPaltiShiuVafa2018,AgrawalObied2018}. Then
\begin{equation}
        \eta_V=-\frac{|m^2|\Mpl^2}{V_0},
\end{equation}
and $|m^2|=3|\eta_V|\Ode H_0^2$. The linearized homogeneous equation is
\begin{equation}
        \ddot\phi+3H_0\dot\phi-3|\eta_V|\Ode H_0^2\phi=0.
\end{equation}
Taking $H\simeq H_0$ over the local interval, solutions behave as $\phi\propto e^{\lambda H_0t}$ with
\begin{equation}
        \lambda_\pm(|\eta_V|)=\frac{-3\pm\sqrt{9+12\Ode |\eta_V|}}{2}.
        \label{eq:lambda-pm}
\end{equation}
The growing mode has $\lambda_+>0$, so generically
\begin{equation}
        x=\frac{\dot\phi}{H_0\Mpl}\simeq \lambda_+\frac{\phi}{\Mpl}.
\end{equation}
The clock bound implies
\begin{equation}
        \frac{|\phi_0|}{\Mpl}\lesssim \frac{x_{\rm max}}{\lambda_+(|\eta_V|)}.
        \label{eq:hilltop-exact-bound}
\end{equation}
For $|\eta_V|\ll1$, expanding \cref{eq:lambda-pm} gives $\lambda_+(|\eta_V|)=\Ode |\eta_V|+O(\eta_V^2),$ so \cref{eq:hilltop-exact-bound} becomes
\begin{equation}
        \frac{|\phi_0|}{\Mpl}\lesssim \frac{x_{\rm max}}{\Ode |\eta_V|}.
\end{equation}
Equivalently near the hilltop, we get
\begin{equation}
        \lambda_V\simeq |\eta_V|\frac{|\phi_0|}{\Mpl},
\end{equation}
\begin{equation}
        x\sim \Ode \lambda_V\sim \Ode |\eta_V|\frac{|\phi_0|}{\Mpl}.
\end{equation}
Thus an $\mathcal{O}(1)$ tachyonic Hessian is compatible with our representative clock bound only if the scalar is currently very close to the hilltop. Alternatively, it is possible to change the initial conditions to suppress the growing mode, but that is dynamical fine tuning rather than a generic allowed region. 

\bibliographystyle{unsrt}
\bibliography{bibliography}

\end{document}